\documentclass{aa}
\begin{document}
\outer\def\gtae {$\buildrel {\lower3pt\hbox{$>$}} \over 
{\lower2pt\hbox{$\sim$}} $}
\outer\def\ltae {$\buildrel {\lower3pt\hbox{$<$}} \over 
{\lower2pt\hbox{$\sim$}} $}
\newcommand{\ergscm} {ergs s$^{-1}$ cm$^{-2}$}
\newcommand{\ergss} {ergs s$^{-1}$}
\newcommand{\ergsd} {ergs s$^{-1}$ $d^{2}_{100}$}
\newcommand{\pcmsq} {cm$^{-2}$}
\newcommand{\ros} {\sl ROSAT}
\newcommand{\exo} {\sl EXOSAT}
\def\rchi{{${\chi}_{\nu}^{2}$}}
\newcommand{\Msun} {$M_{\odot}$}
\newcommand{\Mwd} {$M_{wd}$}
\def\Mdot{\hbox{$\dot M$}}

%
   \title{First {\sl XMM-Newton} observations of a Cataclysmic
   Variable II: the X-ray spectrum of OY Car\thanks{Based on observations
   obtained with XMM-Newton, an ESA science mission with instruments
   and contributions directly funded by ESA Member States and the USA
   (NASA).}}

   \author{Gavin Ramsay\inst{1}, 
France C\'{o}rdova\inst{2}, Jean Cottam\inst{3}, Keith
   Mason\inst{1}, Rudi Much\inst{4}, 
Julian Osborne\inst{5}, Dirk Pandel\inst{2}, Tracey Poole\inst{1},
Peter Wheatley\inst{5}}

   \offprints{Gavin Ramsay: gtbr@mssl.ucl.ac.uk}

\institute{$^{1}$Mullard Space Science Lab, University College London,
Holmbury St. Mary, Dorking, Surrey, RH5 6NT, UK\\
$^{2}$Department of Physics, University of California, Santa
Barbara, California 93106, USA\\
$^{3}$Columbia Astrophysics Laboratory, Columbia University, 
538 West 120th Street, New York, NY 10027, USA\\
$^{4}$Astrophysics Division, ESTEC, 2200, AG Noordwijk, The
Netherlands\\
$^{5}$Department of Physics \& Astronomy, University of Leicester,
University Road, Leicester, LE1 7RH, UK\\}

\authorrunning{G. Ramsay et al}
\titlerunning{The X-ray spectrum of OY Car}

\date{}

\maketitle

\begin{abstract}

We present {\sl XMM-Newton} X-ray spectra of the disc accreting
cataclysmic variable OY Car, which were obtained during the
performance verification phase of the mission. These data were taken 4
days after a short outburst. In the EPIC spectra we find strong iron
K$\alpha$ emission with weaker iron K$\beta$ emission together with
silicon and sulphur lines. The spectra are best fitted with a three
temperature plasma model with a partial covering absorber. Multiple
temperature emission is confirmed by the emission lines seen in the
RGS spectrum and the H/He like intensity ratio for iron and sulphur
which imply temperatures of $\sim$7keV and $\sim$3keV respectively.


\end{abstract}

\section{Introduction}

Cataclysmic variables (CVs) are close binary systems in which the
secondary (usually a dwarf main sequence star) fills its Roche lobe
and transfers material onto the white dwarf primary. In non-magnetic
systems, this material forms an accretion disc around the primary.
Some of these systems show `dwarf nova' outbursts on the timescale of
weeks to months when the system brightens by several magnitudes
lasting a few days to months. During an outburst, the material close
to the white dwarf is optically thick, while in quiescence the gas is
optically thin (Pringle \& Savonije 1979, Narayan \& Popham 1993,
Popham \& Narayan 1995).

X-ray observations of dwarf novae are hampered by their relatively low
flux levels. With observations using large effective areas, such as
{\sl XMM-Newton} (Jansen et al, 2001), we can obtain high signal to
noise X-ray spectra of dwarf novae for the first time. In this paper
we present and analyse X-ray spectra of the dwarf nova OY Car.  In a
companion paper (Ramsay et al 2001), we present and analyse the light
curves of OY Car.

\section{Observations}

Ramsay et al (2001) describe the {\sl XMM-Newton}
observations. Briefly, OY Car was observed twice using {\sl
XMM-Newton}. The first observation (29--30 June 2000) was longer
($\sim$50 ksec in duration for the EPIC detectors) and had a higher
count rate than the second observation (7 August 2000). This paper
concentrates on the first observation which was made 4 days after a
short outburst of OY Car.

The EPIC exposures were taken in full window mode using the medium
filter. The particle background in both the EPIC detectors
(0.1--12keV) (Turner et al 2001) and the RGS (0.3--2.1keV) (den Herder
et al 2001) increased significantly towards the end of the
observation: the high background data were therefore removed from the
analysis. Before extracting spectra of OY Car, the data were processed
using the {\sl XMM-Newton} Science Analysis System released on 2000
July 12.

\section{The RGS spectrum}

\begin{figure*}
\begin{center}
\setlength{\unitlength}{1cm}
\begin{picture}(12,4)
\put(-5.,14.8){\includegraphics{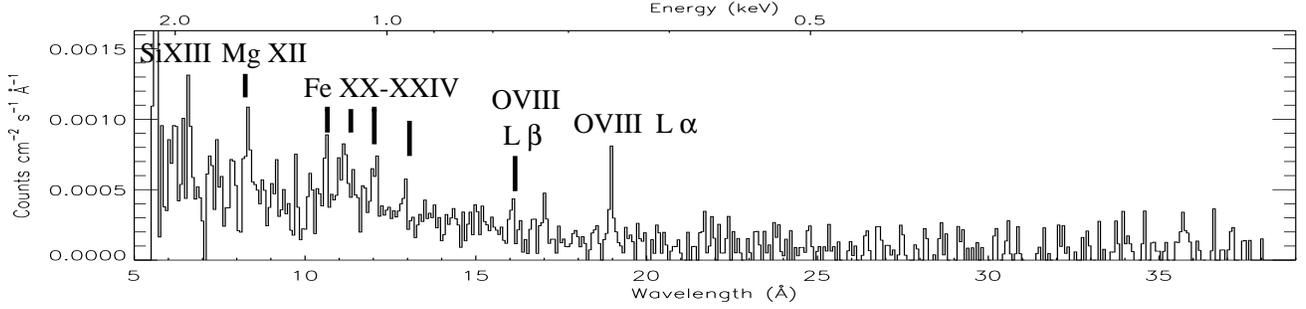}}
\end{picture}
\end{center}
\caption{The background subtracted RGS spectrum taken on 29 June
2000. The exposure was 55 ksec and includes both RGS1 and RGS2 data.}
\label{rgs_spec} 
\end{figure*}

Although the mean background subtracted count rate was low ($\sim 0.05
{\rm ct \, s^{-1}}$) we were able to extract the RGS spectrum using
data from both RGS1 and RGS2 (Figure \ref{rgs_spec}). Prominent
emission lines are seen at 0.654 keV from O VIII Ly $\alpha$ and at
1.473 keV from Mg XII Ly $\alpha$. Fainter lines are seen from O VIII
Ly$\beta$ at 0.775 keV and L-shell transitions of Fe XXIV to Fe XX
between 1.16 keV and 0.96 keV. There is a possible detection of Si
XIII He-like series around 1.854 keV. 

Given the statistical quality of the RGS data we are unable to perform
detailed spectral analysis.  We can, however, use the iron line
detections to roughly constrain the temperature of the plasma assuming
that it is in collisional equilibrium.  The detection of Fe XX lines
requires a plasma temperature between 0.5 and 1.2 keV.  However, the
lack of emission from lower charge-states suggests that the
temperature is actually higher than $\sim$0.8 keV.  The simultaneous
detection of Fe XXIV lines requires an additional temperature
component, since these lines become prominent at temperatures larger
than $\sim$1.3 keV. The lack of observable emission from the O VII
He-like series but the detection of strong O VIII lines, puts a
lower-limit of $\sim$0.4 keV on the oxygen emitting regions.  The RGS
spectrum suggests emission from a distribution of temperatures.

\section{The EPIC spectra}
\label{spectra}

\subsection{General features}

Spectra were extracted from all 3 EPIC cameras using apertures
$\sim30^{''}$ in radius centered on OY Car, chosen so that the
aperture did not cover more than one CCD. This encompasses $\sim$90
percent of the integrated PSF (Aschenbach et al 2000). Background
spectra were extracted from the same CCD on which the source was
detected, scaled and subtracted from the source spectra.

Since the response of the detectors is not well calibrated at present
below $\sim$0.3keV, energies below this were ignored in the following
analysis.  In our fits we used the response file
mos1$\_$medium$\_$all$\_$v3.14$\_$15$\_$tel2.rsp for the MOS detectors
and epn$\_$fs$\_$Y9$\_$medium.rmf for the PN detector. Since this PN
response assumes only single pixel events, we extracted only these
events to make our PN spectrum.

\begin{figure*}
\begin{center}
\setlength{\unitlength}{1cm}
\begin{picture}(13,10.5)
\put(20.5,30.5){\includegraphics{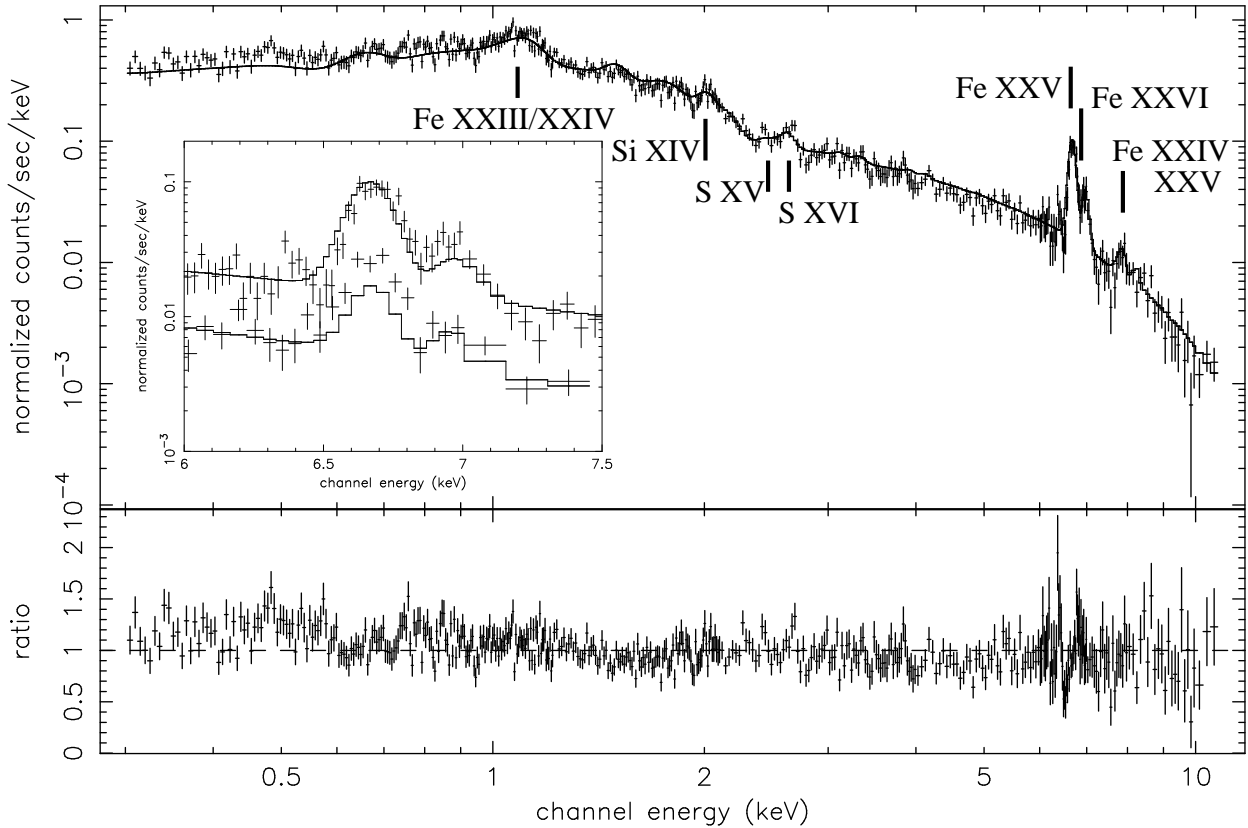}}
\put(-8.7,3.2){\includegraphics{xmm03_fig2b.ps}}
\end{picture}
\end{center}
\caption{Top panel: The integrated EPIC PN spectrum together with the
best fit model using a neutral absorber and a three temperature
thermal plasma model. Bottom panel: a plot of the data/model
ratio. The inset shows the EPIC PN and MOS1 spectra with the best
model fit covering 6.0--7.5keV.}
\label{pnspec} 
\end{figure*}

We show in Figure \ref{pnspec} the integrated EPIC PN spectrum. Strong
iron K$\alpha$ emission lines are seen at $\sim$6.70 \& 6.94keV,
weaker iron K$\beta$ line at 7.90 keV and emission from Fe-L lines
around 1.1keV. Lines are also seen from Si and S. Similar lines are
found in the spectra of the disc accreting CV SS Cyg (Done \& Osborne
1997) and the weakly magnetic CV EX Hya (Ishida, Mukai \& Osborne
1994).

\subsection{Spectral fitting}

The X-ray satellite {\sl ASCA} (0.6--10keV) combined good spectral
resolution together with reasonably large effective area.  Around 15
dwarf novae were observed using {\sl ASCA}.  Amongst the brightest was
SS Cyg. Done \& Osborne (1997) showed that in outburst a
multi-temperature model was needed, while in quiescence a
single-temperature model gave good fits to the data. We therefore
fitted the integrated spectra from the three EPIC cameras
simultaneously using models of increasing complexity.

A single temperature thermal plasma model with a model for absorption
by neutral gas (the {\tt MEKAL} and {\tt WABS} models in {\tt XSPEC})
gave a fit with \rchi=2.23 (890 d.o.f.)  and a best fit temperature of
$kT$=6.1keV. Adding a second thermal plasma model yielded a
significantly better fit: \rchi=2.06 (888 d.o.f.)  ($kT$=3.3,
8.0keV). The normalisation of the various emission components was
allowed to vary between the three EPIC spectra although their relative
normalisations were held the same in each spectrum. Adding a third
plasma model gave \rchi=1.87 (886 d.o.f.)  ($kT$=0.8, 3.3,
7.7keV). Using an F-test this model is better than the previous with a
significance greater than 99.99 percent. The addition of yet another
thermal plasma component did not give a significantly better fit nor
did the addition of a blackbody component.

We note the lack of line emission at 6.4keV. A line at 6.4keV has been
seen in some dwarf novae (eg SS Cyg observed using {\sl ASCA}: Done \&
Osborne 1997) and is due to fluorescence either from the surface of
the white dwarf or surrounding local material. By adding a Gaussian
(fixed at 6.4keV) to the model described above, we derive an upper
limit on the equivalent width of 24eV for a line of width 100eV and
29eV for a line of width 50eV. The fact that we did not detect a
fluorescent line is possibly due to the high inclination of the system
which may prevent us observing any reflection from the accretion disc.

For the sake of clarity, we show in Figure \ref{pnspec} only the EPIC
PN spectrum in full, together with the three temperature thermal
plasma model and the residuals. We also show in figure \ref{pnspec}
the EPIC PN and MOS1 spectra around 6.7keV. Table \ref{fits} lists the
best fit parameters. The need for a multi-temperature model is
consistent with the RGS spectrum which showed emission from plasma
with more than one temperature.

In their observations of SS Cyg in quiescence Done \& Osborne (1997)
found evidence that the absorption component is more complex than a
simple neutral absorber.  We therefore investigated if our X-ray
spectra could be better modelled using a complex absorber. (In this
study we only consider the EPIC PN data because of software problems
in fitting all three EPIC spectra with complex absorbers).  When we
fitted the EPIC PN spectrum alone we found that a two temperature {\tt
MEKAL} model with neutral absorption gave as good fits as a three
temperature {\tt MEKAL} model (\rchi=1.71, 476 d.o.f): a third
temperature component is not necessary. Using a warm absorber (of the
sort described by Cropper, Ramsay \& Wu 1998) vary and fixing the
column density of the neutral absorber to be $N_{\rm H}=3\times10^{19}$
\pcmsq (Mauche \& Raymond 2000) we obtain \rchi=1.61 (475 d.o.f) with
temperatures $kT$=1.2 \& 6.0keV and a warm absorber of temperature
T$\sim5\times10^{5}$ K and column density $N_{\rm H}=1.3\times10^{21}$
\pcmsq. The implied value for the ionisation parameter is $\xi=0.058,
(=L/nr^{2}$ where $L$ is the luminosity, $n$ the density and $r$ the
distance of the absorber from the irradiating source). The addition of
the warm absorber produces significant absorption edges near
$\sim$0.6keV. The signal to noise of the RGS spectra is too low to detect
these edges.  This warm absorber model is significantly ($>$ 99.99 per
cent using the F-test) better than the model which assumed only
neutral absorption. Adding a third {\tt MEKAL} model gave a fit
(\rchi=1.59, 473 d.o.f) which is better than a two temperature model
at the 97.7 per cent level.  We also used a partial covering model
instead of a warm absorber. Using a two-temperature model gives a fit
of \rchi=1.63 (475 d.o.f). A three temperature model gave a
significantly better fit (\rchi=1.54, 473 d.o.f) with temperatures
($kT$=0.7, 2.5, 7.2keV) and a partial covering model of column
$N_{\rm H}=6.3\times10^{21}$ \pcmsq with a covering fraction of 0.51.

We conclude that there is evidence for complex absorption in the X-ray
spectrum of OY Car. We caution, however, that we have used the
integrated spectrum in our study: Ramsay et al (2001) found evidence
for a spectral variation in the light curves. It is left to a future
paper to investigate this.

\subsection{Line studies}

\begin{figure}
\begin{center}
\setlength{\unitlength}{1cm}
\begin{picture}(8,6)
\put(-1.5,-28.5){\includegraphics{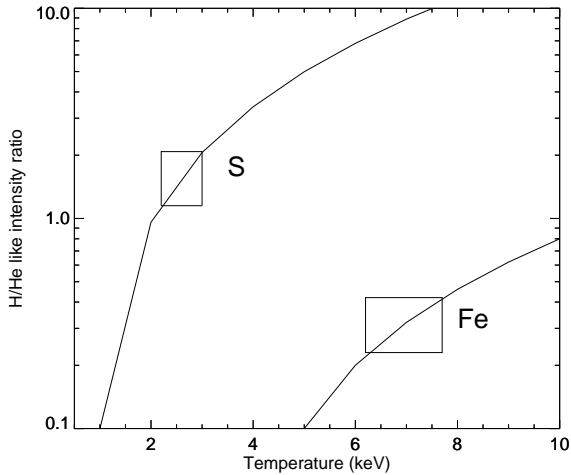}}
\end{picture}
\end{center}
\caption{The H/He like intensity ratio for sulphur and iron K$\alpha$
as a function of temperature. We plot the observed range determined
using the EPIC PN spectrum.}
\label{ion_temp} 
\end{figure}

We can also determine the temperature of the plasma by measuring the
intensity ratio of the H and He like line emission. Using the EPIC PN
spectrum we can determine this ratio for Fe and S. We fitted a thermal
bremsstrahlung emission component and added Gaussian components to fit
the He and H like emission lines. For Fe at 6.70 and 6.95keV we find a
H/He like intensity ratio of 0.31$^{+0.11}_{-0.08}$ and for S at 2.43
and 2.65 keV we find a ratio of 1.15$^{+0.87}_{-0.06}$. Using the {\tt
MEKAL} thermal plasma model in {\tt XSPEC} we obtained the H/He like
intensity ratio verses ionisation temperature for the metal abundance
derived in the fit (table \ref{fits}) and plot this in figure
\ref{ion_temp}.  We find these ratios give temperatures of 6--8keV for
iron K$\alpha$ and 2--3keV for sulphur. Thus, as in our fits using
multi-temperature {\tt MEKAL} models, we need a range of temperatures
to model the data. Indeed the temperatures derived from the line fits
are consistent with the temperatures of the hotter two thermal plasma
components.

\begin{table}
\begin{tabular}{lr}
\hline 
$N_{\rm H}$ (cm$^{-2}$)& $4.8^{+0.5}_{-0.4}\times 10^{21}$\\
Temperature (keV)& $7.7^{+0.5}_{-0.4}$, 3.3$^{+0.2}_{-0.3}$,\\ &
0.76$^{+0.04}_{-0.02}$\\ Metal abundance (solar)& 1.31$^{+0.06}_{-0.08}$\\
Observed flux 0.2--10keV & $4.0\pm0.6 \times 10^{-12}$ \\ 
 (\ergscm)& \\
Bolometric flux (\ergscm) & $5.0\pm1.0\times 10^{-12}$ \\ 
\hline
\end{tabular}
\caption{The best fit parameters for a simultaneous fit to all 3
integrated EPIC spectra using a 3 temperature {\tt MEKAL} thermal 
plasma model.} 
\label{fits}
\end{table}

\subsection{The luminosity}

The flux values shown in Table \ref{fits} are the mean values
determined from each EPIC spectrum. For a distance of 82$\pm$12 pc
(Wood et al 1989) we determine the bolometric X-ray luminosity to be
4.0$\pm0.8\times10^{30}$ \ergss. This luminosity is similar to that
reported for other non-magnetic CVs in quiescence: Pratt et al (1999)
found $L_{\rm X,bol}\sim10^{30}$ \ergss for OY Car using {\sl ROSAT}
data. The luminosity in the {\sl ROSAT} energy band (0.1--2keV) is
$1.9\times10^{30}$ \ergss. We can determine the accretion rate, \Mdot,
from $L_{\rm acc}=GM_{1}\Mdot/R_{1}$, where $L_{\rm acc}$ is the accretion
luminosity and $L_{\rm bl}=L_{\rm acc}/8$ (Popham \& Narayan 1995), where
$L_{\rm bl}$ is the boundary layer luminosity. For a white dwarf mass of
1.0 \Msun (Ramsay et al 2001) and assuming the X-ray flux originates
mainly from the boundary layer, we find a mass accretion rate during
quiescence of 1.3$\times10^{14}$ g s$^{-1}$ or 1.9$\times10^{-12}$
\Msun yr$^{-1}$. The true luminosity and accretion rate maybe
significantly higher since van Teeseling, Beuermann \& Verbunt (1996)
showed that the observed flux is anti-correlated with inclination: for
high inclination systems most of the flux is absorbed by the accretion
disc.

\section{Summary}

We have examined the X-ray spectra of OY Car obtained using {\sl
XMM-Newton}. These spectra have a much higher signal to noise and
spectral resolution than previous X-ray spectra of dwarf nova.
We find strong emission lines of various ionisation species. In
fitting the EPIC spectra we require a multi-temperature plasma. This
is confirmed by the line species seen in the RGS spectrum and also
from the H/He-like intensity ratios of iron and sulphur. Adding a more
complex absorber such as a warm absorber or a partial covering model
to a neutral absorber significantly improves the fit to the data.

\end{document}